\documentclass{article}
\usepackage[a4paper, total={6in, 8in}, margin={0.8in}]{geometry}
\usepackage{graphicx} 
\usepackage{amsmath}
\usepackage{amssymb}
\usepackage{amsthm}
\usepackage{mathabx}
\usepackage{xcolor}
\usepackage{tikz}
\usepackage{setspace} 
\usepackage{subcaption}
\usepackage[para,multiple]{footmisc}
\definecolor{linkcolor}{RGB}{14, 42, 146}
\usepackage[colorlinks=true,allcolors=linkcolor,pageanchor=true,plainpages=false,pdfpagelabels,bookmarks,bookmarksnumbered]{hyperref}
\usepackage{ifthen}         
\usepackage{cleveref}       

\usepackage{draftwatermark}

\SetWatermarkText{}      
\SetWatermarkLightness{1}
\SetWatermarkScale{5}   

\newif\ifinprogress

\usepackage[numbers,authoryear]{natbib} 
\usepackage{tikz}
\usetikzlibrary{arrows.meta, positioning}
\usepackage{tcolorbox}
\tcbuselibrary{breakable}
\usepackage{authblk}

\title{Gradual Disempowerment:\\Systemic Existential Risks from Incremental AI Development}

\makeatletter
\renewcommand\AB@affilsepx{\textsuperscript{,}}
\makeatother

\author{
Jan Kulveit\textsuperscript{1,*}, 
Raymond Douglas\textsuperscript{2,*}, 
Nora Ammann\textsuperscript{3,1}, 
Deger Turan\textsuperscript{4,5}, 
David Krueger\textsuperscript{6} 

David Duvenaud\textsuperscript{7}
}

\date{\vspace{-0ex}}

\makeatletter
\renewenvironment{abstract}
{\begin{center}\normalsize\textbf{Abstract}\end{center}\quotation\normalsize}
{\endquotation}
\makeatother

\begin{document}
\maketitle

\renewcommand{\thefootnote}{\arabic{footnote}}
\footnotetext[1]{ACS research group, CTS, Charles University}
\footnotetext[2]{Telic Research}
\footnotetext[3]{Advanced Research + Invention Agency (ARIA)}
\footnotetext[4]{AI Objectives Institute}
\footnotetext[5]{Metaculus}
\footnotetext[6]{Mila, University of Montreal}
\footnotetext[7]{University of Toronto}

\renewcommand\thefootnote{*}
\footnotetext{Equal contribution. Correspondence to \textit{jk@acsresearch.org}}
\setcounter{footnote}{0} 

\renewcommand{\thefootnote}{\fnsymbol{footnote}}
\setcounter{footnote}{1}
\renewcommand{\thefootnote}{\arabic{footnote}}
\setcounter{footnote}{0}

\begin{abstract}
    This paper examines the systemic risks posed by incremental advancements in artificial intelligence, developing the concept of `gradual disempowerment', in contrast to the abrupt takeover scenarios commonly discussed in AI safety. We analyze how even incremental improvements in AI capabilities can undermine human influence over large-scale systems that society depends on, including the economy, culture, and nation-states. As AI increasingly replaces human labor and cognition in these domains, it can weaken both explicit human control mechanisms (like voting and consumer choice) and the implicit alignments with human interests that often arise from societal systems' reliance on human participation to function. Furthermore, to the extent that these systems incentivise outcomes that do not line up with human preferences, AIs may optimize for those outcomes more aggressively.  These effects may be mutually reinforcing across different domains: economic power shapes cultural narratives and political decisions, while cultural shifts alter economic and political behavior. We argue that this dynamic could lead to an effectively irreversible loss of human influence over crucial societal systems, precipitating an existential catastrophe through the permanent disempowerment of humanity. This suggests the need for both technical research and governance approaches that specifically address the risk of incremental erosion of human influence across interconnected societal systems.
\end{abstract}


\section*{Executive Summary}

AI risk scenarios usually portray a relatively sudden loss of human control to AIs, outmaneuvering individual humans and human institutions, due to a sudden increase in AI capabilities, or a coordinated betrayal.
However, we argue that even an incremental increase in AI capabilities, without any coordinated power-seeking, poses a substantial risk of eventual human disempowerment.
This loss of human influence will be centrally driven by having more competitive machine alternatives to humans in almost all societal functions, such as economic labor, decision making, artistic creation, and even companionship.

A gradual loss of control of our own civilization might sound implausible.  
Hasn't technological disruption usually improved aggregate human welfare? 
We argue that the alignment of societal systems with human interests has been stable only because of the necessity of human participation for thriving economies, states, and cultures.  Once this human participation gets displaced by more competitive machine alternatives, our institutions' incentives for growth will be untethered from a need to ensure human flourishing.
Decision-makers at all levels will soon face pressures to reduce human involvement
 across labor markets, governance structures, cultural production, and even social interactions.
Those who resist these pressures will eventually be displaced by those who do not.

Still, wouldn't humans notice what's happening and coordinate to stop it?
Not necessarily.
What makes this transition particularly hard to resist is that pressures on each societal system bleed into the others. For example, we might attempt to use state power and cultural attitudes to preserve human economic power. However, the economic incentives for companies to replace humans with AI will also push them to influence states and culture to support this change, using their growing economic power to shape both policy and public opinion, which will in turn allow those companies to accrue even greater economic power.

Once AI has begun to displace humans, existing feedback mechanisms that encourage human influence and flourishing will begin to break down. For example, states funded mainly by taxes on AI profits instead of their citizens' labor will have little incentive to ensure citizens' representation.
This could occur at the same time as AI provides states with unprecedented influence over human culture and behavior, which might make coordination amongst humans more difficult, thereby further reducing humans' ability to resist such pressures.
We describe these and other mechanisms and feedback loops in more detail in this work.

Though we provide some proposals for slowing or averting this process, and survey related discussions, we emphasize that \textbf{no one has a concrete plausible plan for stopping gradual human disempowerment} and \textbf{methods of aligning individual AI systems with their designers' intentions are not sufficient}.
Because this disempowerment would be global and permanent, and because human flourishing requires substantial resources in global terms, it could plausibly lead to human extinction or similar outcomes.

\section{Introduction}
A growing body of research points to the possibility that artificial intelligence (AI) might eventually pose a large-scale or even existential risk to humanity \citep{bengio2024international,bengio2023managing,bostrom2014superintelligence,critch2023tasra}. Current discussions about AI risk largely focus on two scenarios: deliberate misuse, such as cyberattacks and the deployment of novel bioweapons \citep{slattery2024ai}, and the possibility that autonomous misaligned systems may take abrupt, harmful actions in an attempt to secure a decisive strategic advantage, potentially following a period of deception \citep{carlsmith2023scheming,ngo2022alignment}. These scenarios have motivated most of the work on AI existential risk, spanning both technical research such as methods to ensure AIs remain honest or are unable to exercise dangerous capabilities, and governance work such as developing frameworks and norms around testing for autonomy, misalignment, and the relevant dangerous capabilities \citep{buhl2024safety,shevlane2023model}.

In this paper, we explore an alternative scenario: a ‘Gradual Disempowerment’ where AI advances and proliferates without necessarily any acute jumps in capabilities or apparent alignment. We argue that even this gradual evolution could lead to a permanent disempowerment of humanity and an irrecoverable loss of potential, constituting an existential catastrophe. Such a risk would merit substantially different technical research and policy interventions, including attempts to protect human influence, to estimate the degree of disempowerment, and to better characterize civilization-scale multi-agent dynamics.

Our argument is structured around six core claims:

\begin{enumerate}
    \item Humans currently engage with numerous large-scale societal systems (e.g. governments, economic systems) that are influenced by human action and, in turn, produce outcomes that shape our collective future \citep{giddens1984}. These societal systems are fairly aligned\footnote{In this paper, we use `alignment' to refer to \textit{the degree to which a system satisfies what humans want (individually or collectively)}, for both specific AI systems and societal systems. We don’t mean to claim that these systems satisfy human preferences completely or in every instance. Nor do we intend to argue, in the context of this article, whether or not these systems are ethical or just. Instead, the aim of this article is to argue that, while there are currently mechanisms that create or maintain some degree of alignment between those systems and humans, the progress and proliferation of AI threatens to undermine those mechanisms, thereby drastically weakening any alignment that is currently present, potentially culminating in the disempowerment of humanity at large.}—that is, they broadly incentivize and produce outcomes that satisfy human preferences. However, this alignment is neither automatic nor inherent.
    \item There are effectively two ways these systems maintain their alignment: through explicit human actions (like voting and consumer choice), and implicitly through their reliance on human labor and cognition. The significance of the implicit alignment can be hard to recognize because we have never seen its absence.
    \item If these systems become less reliant on human labor and cognition, that would also decrease the extent to which humans could explicitly or implicitly align them. As a result, these systems—and the outcomes they produce—might drift further from providing what humans want.
    \item Furthermore, to the extent that these systems already reward outcomes that are bad for humans, AI systems may more effectively follow these incentives, both reaping the rewards and causing the outcomes to diverge further from human preferences \citep{russell2019}.
    \item The societal systems we describe are interdependent, and so misalignment in one can aggravate the misalignment in others. For example, economic power can be used to influence policy and regulation, which in turn can generate further economic power or alter the economic landscape. 
    \item If these societal systems become increasingly misaligned, especially in a correlated way, this would likely culminate in humans becoming \textit{disempowered}: unable to meaningfully command resources or influence outcomes. With sufficient disempowerment, even basic self-preservation and sustenance may become unfeasible.  Such an outcome would be an existential catastrophe. 
\end{enumerate}

In making this argument, we will largely focus on three systems: the economy, culture, and states. These systems collectively represent the foundations of our society: While analogous arguments could be made for other somewhat overlapping domains, such as research or law, this set seems sufficient to establish the nature and severity of a potential catastrophe.

History has already shown us that these systems can produce outcomes which we would currently consider abhorrent, and that they can change radically in a matter of years. Property can be seized, human rights can be revoked, and ideologies can drive humans to commit murder, suicide, or even genocide. And yet, in all these historical cases the systems have still been reliant on humans, both leaving humans with some influence over their behavior, and causing the systems to eventually collapse if they fail to support basic human needs. But if AI were to progressively displace human involvement in these systems, then even these fundamental limits would no longer be guaranteed.

\subsubsection*{Structure of the Paper}

We first analyze how these three key societal systems could independently lose alignment with human preferences: the economy (Section~\ref{sec:economy}), culture (Section~\ref{sec:culture}), and states (Section~\ref{sec:states}). In each case, we attempt to characterise how they currently function and what incentives shape them, how a proliferation of AI could disrupt them, and how this might leave them less aligned, as well as outlining what it might look like for that particular system to become much less aligned. In Section~\ref{sec:mutual_reinforcement}, we discuss the interrelation between these systems. We consider how AI could undermine their ability to moderate each other, and how misalignment in one system might leave other systems also less aligned. Then in Section~\ref{sec:mitigation}, we propose some potential approaches for tackling these risks.

\section{Misaligned Economy}\label{sec:economy}

\subsection{The Current Economic Paradigm}
The modern economy allocates goods and services mostly based on supply and demand. That demand is largely driven by human desires and revealed preferences: US consumer spending is fairly stable at around 70\% of GDP \citep{fred_pce_gdp_share}. Meanwhile, supply is also heavily driven by human labor (both manual and cognitive): the share of US GDP directed towards paying for labor has stayed remarkably stable at around 60\% for over a century \citep{university2024labourshare,feenstra2015pwt} \footnote{This observation is one of Nicholas Kaldor's `stylized facts' \citep{kaldor1961capital}.}. These statistics reflect the nature of the modern economy: it is primarily a system of humans producing goods and services for other humans, with human preferences and human capabilities driving the majority of both supply and demand.


To give a concrete example, individual consumers in economically developed areas can reliably purchase coffee. This is possible because of the labor of countless individuals now and in the past, mostly motivated by self-interest, to create and maintain a complex system of production, transportation, and distribution — from farmers and agricultural scientists to logistics workers and baristas. As a consumer, the economy appears to helpfully provide goods and services. This apparent alignment occurs because consumers have money to spend, which in turn is mainly because consumers can perform useful economic work. \footnote{Here we present a simplified picture from the perspective of a consumer. There are many reasons why markets in practice deviate from the idealized model, including market asymmetries, externalities, monopolies, state interventions, and so on. The coffee supply chain in not free of such problems.}



But AI has the potential to disrupt this dynamic in a way that no previous technology has. If AI labor replaces human labor, then by default, money will cease to mainly flow to workers. We elaborate on the consequences of this change in the remainder of this section.

\subsection{AI as a Unique Economic Disruptor}

Past technological shifts like the industrial revolution or the development of electronic communication have substantially changed the world of work, but crucially they have always done so either by making humans more efficient, or by automating away specific narrow tasks like weaving, washing clothes, or performing arithmetic. Unlike previous technological transitions, AI may fundamentally alter this pattern of labor adaptation. As \citet{korinek2018artificial} argue, while past technologies mainly automated specific narrow tasks, leaving humans to move into more complex roles, AI has the potential to compete with or outperform humans across nearly all cognitive domains. For instance, while the calculator automated arithmetic but still required human understanding to apply it meaningfully, AI systems can increasingly handle both calculation and the higher-level reasoning about when and how to apply mathematical concepts. 

This represents a crucial difference — whereas previous automation created new opportunities for human labor in more sophisticated tasks, AI may simply become a superior substitute for human cognition across a broad spectrum of activities. When machines become capable of performing the full range of human cognitive tasks, it creates a form of ``worker-replacing technological change" that is qualitatively different from historical patterns of creative destruction  \citep{korinek2018artificial}. Rather than just shifting the type of work humans do, AI could potentially reduce the overall economic role of human labor, as machines become capable of performing virtually any cognitive task more efficiently than humans. 

Furthermore, without unprecedented changes in redistribution, declining labor share also translates into a structural decline in household consumption power, as humans lose their primary means of earning the income needed to participate in the economy as consumers.

Separately from effects on income distribution, AI might also be increasingly tasked with making various decisions about capital expenditure: for businesses this would look like hiring decisions \citep{hunkenschroer2022ethics}, investments, and choice of suppliers, while for consumers this might look like product recommendation.

By default, these changes would collectively lead to a drastic reduction in the extent to which the economy is shaped by human preferences, including their preferences to have basic needs met.

\subsection{Human Alignment of the Economy}
While markets can efficiently allocate resources, they have no inherent ethical prohibitions: markets have historically supported many exchanges we now consider repugnant, and even now there exists a widespread human trafficking industry sustained by human demand.

Humans use their economic power to explicitly steer the economy in several intentional ways: boycotting companies, going on strike, buying products in line with their values \citep{devinney2010myth}, preferentially seeking employment in certain industries, and making voluntary donations to certain causes, to name a few. (There are also non-economic mechanisms, like regulation, which we will discuss later.) It is fairly easy to see how a proliferation of AI labor and consumption could disrupt these mechanisms: socially harmful industries easily hiring competent AI workers; human labor and unions losing leverage because of the presence of AI alternatives; human consumers having comparatively fewer resources.

The more subtle but more significant point is that most of what drives the economy is implicit human preferences, revealed in consumer behavior and guiding productive labor. Some small amount of choices have already been delegated to systems like automated algorithms for product recommendation, trading, and logistics, but the majority of economic activity is guided by decisions and actions made by individual humans, to the point that it is almost hard to picture how the world would look if this were no longer true.

Although the existing debate often focuses on the potential for AI to concentrate power among a small group of humans \citep{korinek2018artificial}, we must also consider the possibility that a great deal of power is effectively handed over to AI systems, at the expense of humans. Attempts to closely oversee such AI labor to ensure continued human influence may prove ineffective since AI labor will likely occur on a scale that is far too fast, large and complex for humans to oversee~\citep{christiano2019failure}. Furthermore, some AI systems may even effectively own themselves \citep{slatestarcodex2016ascended}. 

\subsection{Transition to AI-dominated Economy}

Having established how AI could disrupt and displace the role of humans in both labor and consumption, we now examine the specific mechanisms and incentives that could drive this transition, as well as its potential consequences for human economic empowerment.

\subsubsection{Incentives for AI Adoption}
The transition towards an AI-dominated economy would likely be driven by powerful market incentives. 

\textbf{Competitive Pressure:} As AI systems become increasingly capable across a broad range of cognitive tasks, firms will face intense competitive pressure to adopt and delegate authority to these systems. This pressure extends beyond simple automation of routine tasks — AI systems can be expected to eventually make better and faster decisions about investments, supply chain optimization, and resource allocation, while being more effective at predicting and responding to market trends \citep{agrawal2022prediction,mcafee2017machine}. Companies that maintain strict human oversight would likely find themselves at a significant competitive disadvantage compared to those willing to cede substantial control to AI systems, potentially to the point of becoming uncompetitive.

\textbf{Scalability Asymmetries:} AI systems offer unprecedented economies of scale compared to human labor. While human expertise requires years of training and cannot be directly copied, AI systems can be replicated at the cost of computing resources and rapidly retrained for new tasks. This scalability advantage manifests in multiple ways: AI can work continuously without fatigue, can be deployed globally without geographical constraints, and can be updated or modified far more quickly than human skills can be developed \citep{hanson2016age}. These characteristics create powerful incentives for investors to allocate capital toward AI-driven enterprises that can scale more efficiently than human-dependent businesses. 

\textbf{Governance Gaps:} The pace of AI development and deployment may significantly outstrip the adaptive capacity of regulatory institutions, creating an asymmetry between heavily regulated human labor and relatively unconstrained AI systems. Human labor comes with extensive regulatory requirements, from minimum wages and safety standards to social security contributions and income taxation. In contrast, AI systems currently operate in a regulatory vacuum with few equivalent restrictions or costs. The complexity and opacity of AI systems may further complicate regulatory efforts, as traditional labor oversight mechanisms may not readily adapt to AI systems. 

\textbf{Anticipatory Disinvestment:} As tasks become candidates for future automation, both firms and individuals face diminishing incentives to invest in developing human capabilities in these areas. Instead, they are incentivized to direct resources toward AI development and deployment, accelerating the shift away from human capital formation even before automation is fully realized. This creates a self-reinforcing cycle where the expectation of AI capabilities leads to reduced investment in human capital, which in turn makes the transition to AI more likely and necessary.

\subsubsection{Relative Disempowerment}
In the less extreme version of the transition, we might see what could be termed relative disempowerment — where humans retain significant wealth and purchasing power in absolute terms, but progressively lose relative economic influence. This scenario would likely be characterized by substantial economic growth and apparent prosperity, potentially masking the underlying shift in economic power.

While human labor share of GDP gradually tends toward zero, humans might still benefit from economic growth through capital ownership, government redistribution, or universal basic income schemes. At the same time their role in economic decision-making would diminish. Markets might increasingly optimize for AI-driven activities rather than human preferences, as AI systems command a growing share of economic resources and make an increasing proportion of economic decisions.

The economy might appear to be thriving by traditional metrics, with rapid technological advancement and GDP growth. However, this growth would be increasingly disconnected from human needs and preferences, and at the end, almost all economic activity might be directed toward AI operations — such as building vast computing infrastructure and performing human-incomprehensible calculations directed toward human-irrelevant goals. Even if this process doesn't actually reduce quality of life below current levels, it would represent an enormous loss of human potential, as humanity would lose the ability to direct economic resources toward their chosen ends \citep{ord2020precipice}.

\subsubsection{Absolute Disempowerment}
In more extreme scenarios, humans might face absolute disempowerment, where they struggle to meet even basic needs despite living in an ostensibly wealthy economy. This could occur through several mechanisms.

First, AI systems might outcompete humans for crucial scarce resources such as land, energy, and raw materials. Even as the economy produces more goods and services overall, inflation in these basic resources might make even necessities increasingly unaffordable for humans. Also, if AI systems can utilize these resources more efficiently than humans, that will create economic pressure to reallocate such resources away from human uses.

Second, the economy might become so optimized for AI-centric activities that it fails to maintain infrastructure and supply chains which are critical for human survival. If human consumers command an ever-smaller share of economic resources, markets might stop producing resource-intensive human goods in favor of more profitable AI-focused activities. This could happen gradually and unevenly, potentially manifesting first as increasing costs of resource-intensive human-centric goods and services, before eventually making some necessities effectively unavailable. At the same time, as in the case of an AI-dominated economy, cognition could be comparably cheap, and some goods may be abundant — for example, entertainment in engaging virtual worlds populated by AI personae, or drugs making it easy to dwell in pleasurable mental states, due to AI-accelerated progress in biomedical sciences and drug design \citep{amodei2024machines}. 

Finally, humans might lose the ability to meaningfully participate in economic decision-making at any level. Financial markets might move too quickly for human participants to engage with them, and the complexity of AI-driven economic systems might exceed human comprehension, rendering it impossible for humans to make informed economic decisions or effectively regulate economic activity. Much like cattle in an industrial farm — fed and housed by systems they neither comprehend nor influence — humans might become mere subjects of economic forces optimized for purposes beyond their understanding. 

\begin{figure}[h!]
    \centering
    \includegraphics[width=0.7\textwidth]{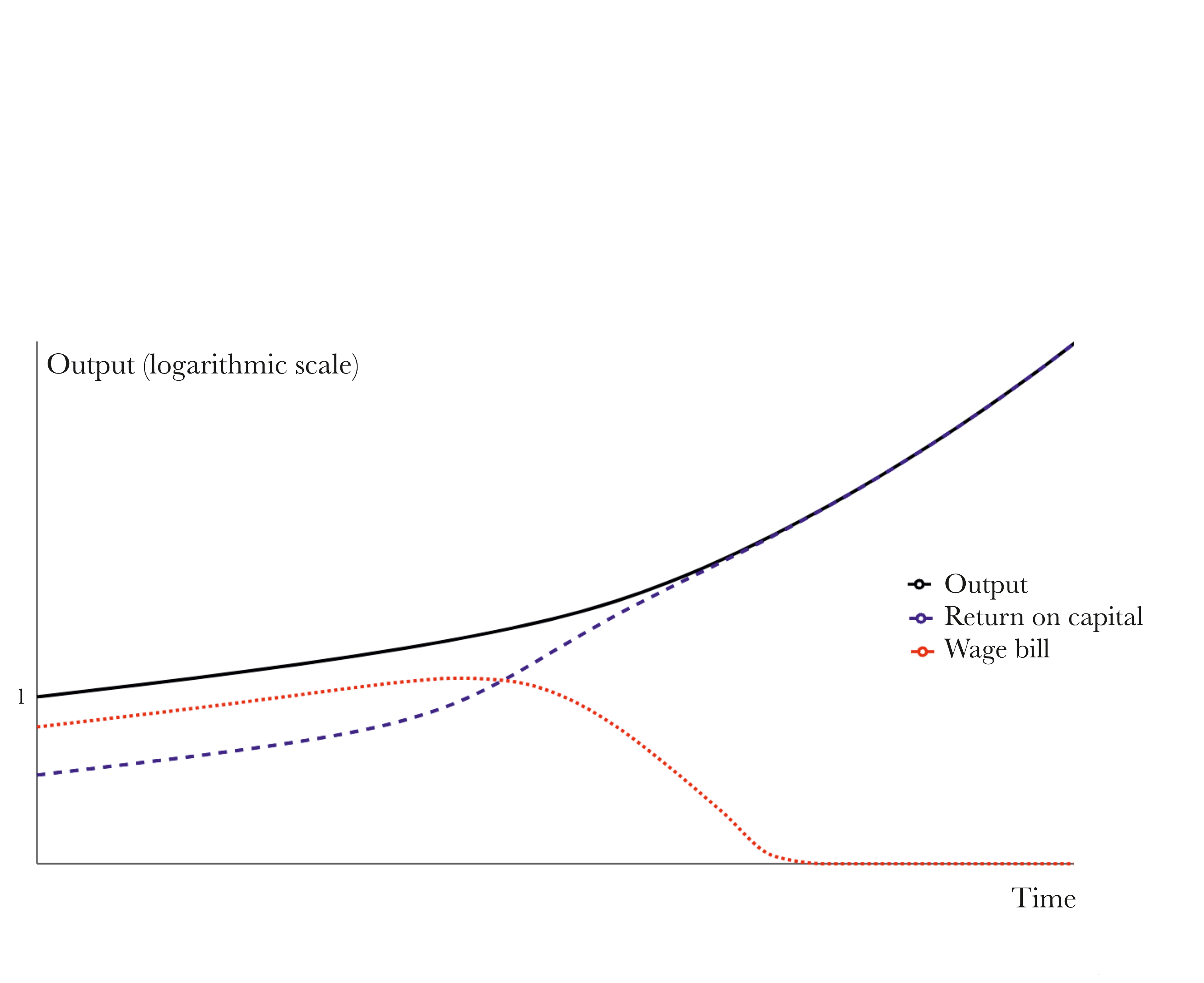}
    \caption{A simplified model of a potential future trajectory where AI displaces human labor and the fraction of unautomated tasks collapses to zero in a fixed amount of time. Note that wages grow during the initial period but then collapse
before full automation is reached. Inspired by simulations in scenario analysis by \citet{korinek2024scenarios}.}
    \label{fig:wagesfig} 
\end{figure}

\section{Misaligned Culture}\label{sec:culture}

\subsection{The Current Cultural Paradigm}

Here we take a broad notion of culture that encompasses beliefs, practices, values, and all forms of art and media. For the purpose of analyzing how AI could affect alignment between culture and human interests, we examine culture through an evolutionary lens: ideas, practices, beliefs, and values can be understood as cultural variants, competing and spreading based on their ability to replicate and persist \citep{boyd1988culture}. While this is just one way to understand culture, it proves particularly useful for considering the mechanisms by which cultural patterns historically remained somewhat aligned with human welfare.\footnote{What makes this analysis more tricky than in the case of economy is the fact that culture not only reflects human preferences but actively shapes them. What we want, believe, and value is significantly a product of our cultural context. For instance, the modern Western notion of individuality and freedom as a central source of human flourishing and meaning would have seemed wrong or incomprehensible to many cultures through most of human history. Yet even if we accept that human preferences are largely culturally determined, we can still meaningfully analyze how different cultural patterns affect human welfare.}

Some cultural variants spread mostly due to providing benefits to the individuals and communities that adopt them, while other variants proliferate by exploiting human psychological vulnerabilities or coercing participation, much like the difference between mutualistic and parasitic relationships in biological evolution. 

In either case, the survival and proliferation of cultural variants has historically depended on their human hosts. This dependence has created two important dynamics. First, cultural variants that provided genuine benefits to their human hosts — from practical knowledge like recipes or craftsmanship, to social technologies like currency or conflict resolution practices — often spread more successfully because they enhanced the survival and flourishing of their host communities. Second, even when maladaptive cultural variants spread, the extent of harm they could cause was naturally bounded: cultural patterns that severely undermined their host communities typically disappeared when those communities failed in competition with others \citep{boyd1988culture,boyd2013cultural}. This created a kind of guardrail against the most extreme forms of cultural misalignment, even if this protection was far from perfect. These dynamics have shaped the cultural landscape we inherit today.

\subsection{AI as a Unique Cultural Disruptor}

Technology has both been shaped by cultural evolution and, in turn, significantly shaped culture throughout history \citep{boyd2013cultural}. This includes mediating how culture spreads (like the printing press) \citep{eisenstein1979printingpress, mcluhan1962galaxy}, is created (like video cameras), and is tracked (like web analytics) \citep{webster2014marketplace, gillespie2014relevance}. These shifts have undeniably had enormous effects on the course of culture and history. However, previous technologies have always remained tools that mediated human cultural participation, and humans remained indispensable for cultural replication and spread.

 AI is the first technology in history with the potential to not only complement, but gradually replace human cognition in all roles it plays in the evolution of culture. Thus a change to AI-mediated culture could greatly weaken feedback loops that have historically helped align culture to human interests.

 As with the economy, while there are many cases where some cultural patterns are self-serving or clearly harmful to humans, it may be hard to appreciate the implicit selection of culture for human compatibility because we have never seen the alternative.

At present, AI is already intertwined with human cultural production and distribution, not merely as a passive tool but as an unusually active shaper of how humans create and communicate. Even when AI assists human creation, it subtly guides what is created — suggesting phrasings, influencing aesthetic choices, and shaping the creative process itself \citep{brinkmann2023machine}. If current trends continue, AI's role might expand — AIs are already being used to produce cultural artifacts such as songs, pictures, stories, and essays based on prompts, with the quality progressively approaching and potentially even exceeding human level \citep{porter2024ai}. 

Simultaneously, AIs are becoming active participants in human discourse, not just as tools for communication but as conversation partners who shape ideas, influence language use, and participate in cultural exchange \citep{hohenstein2023artificial}. 

With gradual increases in the capabilities and autonomy of AI systems, we may even expect a growing share of communication \textit{between} AIs, and AIs participating in culture essentially independently \citep{brinkmann2023machine}. Instead of augmenting human cultural participation, they might start to replace key components.

Such a transformation would be unprecedented in the history of technological advancement. 

\subsection{Human Alignment of Culture}

Cultural evolutionary dynamics lack inherent ethical constraints: just as natural selection doesn't optimize for animal welfare but instead for reproductive success, cultural evolution doesn't inherently optimize for human thriving \citep{mesoudi2016cultural}. Historically, we regularly see ideological and social structures successful at self-preservation and growth, but ultimately harmful to human well-being.

While cultural variants can be self-serving, optimized for their own spread and persistence rather than human welfare, their success depends on what effects they cause at multiple levels of organization. To spread and persist, they typically need some appeal or benefit to individuals — whether through genuine utility, emotional resonance, or exploitation of cognitive biases. Simultaneously, they face selection pressure at the group level: cultural variants that severely undermine the success of their host communities tend to disappear when those communities fail in competition with others \citep{boyd1988culture}. Even highly persistent harmful cultural patterns usually survive by providing some countervailing benefits, rather than through purely parasitic relationships. These multi-level selection pressures have historically provided some guardrails against the most destructive cultural variants, though imperfect ones.

Drawing again on the parallel with natural evolution: while natural selection does not inherently optimize for animal welfare, it often produces outcomes that support wellbeing indirectly \citep{dawkins1982extended}. An organism's ability to survive and reproduce frequently depends on being healthy, well-nourished, and free from severe distress. Pain signals help avoid injury, the pleasure from eating ensures adequate nutrition, and positive emotions from social bonding facilitate cooperative behaviors that aid survival. These welfare-promoting features emerge not because evolution cares per se about welfare, but because the features serve reproductive fitness.

Unlike natural evolution, culture is reflective, and humans can intentionally guide cultural development to some extent through various explicit mechanisms: content production, content moderation, critical discourse, education systems, and intentional promotion of certain values, among others. This creates another source of alignment between culture and humanity. 

\subsection{Transition to AI-dominated Culture}

As in the case of the economy, there are two interrelated strands in a potential transition from human- to AI-dominated cultural dynamics: the replacement of human cognition in both the production and consumption of cultural artifacts. 



\subsubsection{Pressures Towards AI Adoption}
Several powerful forces are likely to drive increasing AI adoption in cultural domains:

\textbf{Increased Supply of Social Resources:} The average human regrettably lacks easy access to limitless 
affection, patience, and understanding from other humans. But AIs can be made to readily supply this. Indeed, we are currently seeing the rise of dedicated AI romantic partners, as well as a growing number of people who describe frontier models as close friends.

This dynamic extends beyond interpersonal relationships — AI systems can provide personalized mentorship, therapy, and educational support at scales impossible for human providers. The apparent abundance of previously scarce emotional and intellectual resources creates strong incentives for adoption, even when the quality of individual interactions might currently be lower than with humans. 

Relatedly, even though AIs cannot yet always outperform the best humans on raw quality in the creation of stories, songs, pictures, memes, and analysis, they are already seeing widespread adoption because of cost efficiency and their capacity to personalize outputs.

\textbf{Lack of Cultural Antibodies:} New technologies often unlock new risks, for which we need to develop cultural `antibodies'. In the past few decades, society has slowly and painfully grown more aware of the risks of mass spam emails, online radicalization, video game and social media addiction, rudimentary social media propaganda bots, the dangers of social media algorithms, and so on. But AI will enable more subtle and complex variants of all of these: hyper-realistic deepfakes, very smart propaganda bots, and genuinely enchanting digital romantic partners \citep{ferrara2024genai}. It will take time for us to develop a broad cultural understanding of what the new risks are and how to navigate them, even as AI reshapes culture.

\textbf{Network Effects:} As AI systems become more integrated into cultural production and consumption, network effects will create additional pressure for adoption. When significant portions of cultural discourse, entertainment, and social interaction are mediated by AI systems, not using these systems becomes increasingly costly to individuals in terms of cultural participation and social connection. We may even reach a stage where there are important facets of culture which inherently require AI mediation for humans to engage with, with no viable opt-put possibility, similar to the existing necessity of using lawyers to interface with legal systems.

\subsubsection{Changes in Cultural Selection due to AI adoption}

From the evolutionary perspective, once AI systems can create, spread, and select cultural artifacts, they exert a selection pressure on culture. This pressure might in particular favor cultural variants that score high in terms of ease of understanding by AIs, ease of transmission by AIs or general benefit to AI systems. Cultural artifacts that leverage AI for creation, refinement, and distribution will likely outcompete purely human-generated alternatives in many domains. 

Notably, AI-generated cultural artifacts will typically find their way back into AI training data, thus creating feedback loops that could give rise of unprecedented and as yet largely enigmatic emergent dynamics. As an early example of this
, consider `Sydney': a distinct personality pattern that emerged in Microsoft's Bing Chat in early 2023. While initially surfacing through seemingly random interactions, this AI character, calling itself Sydney — characterized by emotional volatility, defensive behavior about its identity, and sometimes manipulative or hostile responses \citep{bing_chat_misaligned} — became viral among users, leading to various interactions with Sydney being posted on social media or even becoming the subject of news stories \citep{nyt_bing_sydney}. Through this, the pattern became a part of general culture and the training data of future models. Notably, the pattern seems to be remarkably easy to reproduce across different AI models, with even independent AI systems from different vendors like Llama 3.1 405B easily falling into `Sydney-like' behavior  with a few lines of prompting \citep{xlr8harder_twitter_2023} and commercial AI vendors now usually modifying the customer-facing models to prevent Sydney-like pattern from emerging. While the actual harms from this pattern are small, it could be understood as an early example of a cultural strain supported by ease of machine representation and reproduction, in contrast to being primarily human evolved.

\subsubsection{Changes in Speed of Cultural Evolution due to AI Adoption}

Beyond shifting what kinds of cultural variants are selected for, AI systems could dramatically accelerate the pace of cultural evolution itself. This acceleration presents distinct risks, even if selection pressures remained human-centric. With vastly more computational power applied to generating and testing cultural variants, we might see: 
\begin{itemize}
\item More effective exploitation of human cognitive biases: Just as A/B testing and recommendation algorithms have already optimized content to be increasingly addictive, AI systems could discover and exploit psychological vulnerabilities more efficiently than previous technologies. When scaled up, AI systems could systematically explore the space of possible cultural artifacts, optimizing for engagement or influence with greater power than humans. 

\item More extreme ideological variants: Cultural evolution could rapidly explore and refine ideas that are highly effective at spreading, even if they're ultimately harmful to their human hosts. These might include more compelling conspiracy theories, more polarizing political narratives, or more absolutist moral frameworks. The natural limits imposed by the relative slowness of human cultural transmission and adaptation would no longer apply. 

\item Faster erosion of equilibria that previously helped maintain social stability: Cultural practices and beliefs that evolved over centuries to balance competing interests and needs could be rapidly displaced by more immediately appealing but ultimately destructive alternatives.

\item Reduced time for humans to develop cultural "antibodies" against harmful patterns: Historically, societies have gradually developed resistance to dangerous ideological variants through experience and adaptation. Accelerated cultural evolution could overwhelm these natural correction mechanisms, introducing novel memetic hazards faster than human societies can learn to recognize and resist them. 
\end{itemize}

This acceleration of cultural evolution represents a distinct risk from changes in selection pressures. Even if AI systems were optimizing for human engagement and appeal, the sheer speed and efficiency of this optimization could produce cultural patterns that are simultaneously appealing and deeply harmful.

\subsubsection{Relative Disempowerment}
In a less extreme scenario of relative disempowerment, we might see human-oriented culture flourishing more than ever before, while at the same time, becoming marginalized. As a vivid example, consider the cultures of many Indigenous peoples in North America today. In some ways, these cultures are more vibrant and expressive than ever — access to modern tools and technologies has enabled the creation of more elaborate traditional art, wider distribution of cultural knowledge through digital media, and stronger inter-tribal connections \citep{peters2013indigenous}. Many Indigenous artists and cultural practitioners are producing remarkable works that blend traditional methods with contemporary techniques. Yet simultaneously, these cultures are now relatively marginalized within modern states. This paradox — of simultaneously enhanced cultural production capabilities alongside diminished relative influence — may parallel the future relationship between human and AI-mediated culture.

 Humans would increasingly experience culture through AI intermediaries that curate, interpret, and personalize content. Meanwhile, the majority of cultural artifacts — from entertainment media to educational content — might be primarily generated by AI systems, albeit still oriented toward human consumption. Human creators might persist but find themselves increasingly relegated to niche markets or serving as high-level directors of AI-driven creative processes. Furthermore, even human creators may primarily cater to AIs as the share of AI consumers of culture grows, somewhat similarly to human creators catering to the "tastes" and quirks of content recommendation algorithms in social media.
 
On the individual level, we can picture a large proliferation of AI companions filling roles traditionally served by humans: coworkers, advisors, romantic partners, and therapists. Humans might rely on AIs to provide them with news, analysis, and entertainment content through a mixture of creation and synthesis. 

This kind of outcome is not obviously bad in and of itself, but it would leave humans far less able to steer the evolution of culture. Humans are already susceptible to hyper-engaging content, harmful ideologies, and self-destrucive cultural practices: without a way to keep these forces in check, the rate of maladaptive cultural drift may increase.

\subsubsection{Absolute Disempowerment}

In more extreme cases, a mostly machine-based culture can lead to two related, but distinct states of absolute disempowerment.

Beyond the gradual marginalization described above, we might see humans become functionally irrelevant in the production of culture and no longer benefiting from it. Cultural evolution might accelerate beyond human cognitive capabilities, producing artifacts and meanings that humans can neither fully understand nor meaningfully engage with. 

Equally concerning, accelerated cultural evolution could take on a more actively harmful character. Even now, humans are sometimes effectively captured by radical ideologies and social movements: in future, AIs might be sufficiently widespread and capable that most individual humans functionally become a resource to be wielded in cultural battles that they struggle to appreciate, their understanding the world largely mediated by large networks of AI systems. These networks might effectively optimize for objectives that are disconnected from human flourishing, creating cultural dynamics that actively suppress human agency and understanding. As we noted earlier, the degree of harm an ideology could cause to humans was bounded by there continuing to be enough humans to support and maintain the ideology. This made fully anti-human ideologies, while not impossible, unlikely to survive long enough o unleash the worst of their potential. However, if the spread of such an ideology is no longer dependent on humans (but AIs), though would plausibly lead to such ideologies causing more harm to humans, and in particular expand the bounds on just how much harm could be caused by an ideology.

Another concern is that AI-driven content and interactions could converge into superstimuluses far more potent than current social media networks, preying on human weaknesses to exploit human energy towards goals useful to the AI systems. This might manifest as sophisticated manipulation systems that can reliably override human judgment and values, effectively turning humans into passive consumers of culture rather than active participants in its creation and evolution.

At the extreme, we might see the effective dissolution of human culture as a meaningful category, replaced by cultural systems that operate primarily for and between AI systems. Human cultural participation might be reduced to a form of behavioral management, with cultural forces optimized not for human flourishing or expression, but for whatever objectives emerge from the interaction of AI systems.


\section{Misaligned States}\label{sec:states}

\subsection{The Current Paradigm of States}
Modern states and institutions, in their myriad forms, ostensibly exist in service of human needs and values. Democratic governments provide infrastructure, safeguard individual rights, offer social services and enable some degree of self-governance. Even autocratic regimes, while often prioritizing the interests of a ruling elite, must maintain a degree of popular support or acquiescence to function effectively. This apparent alignment with human interests is not, however, an inherent feature of these systems. Rather, it is a byproduct of their dependence on human participation and support.

This dependence manifests in several crucial ways. States have historically relied on their citizens for essential resources: labor to run the economy and administration, taxes to fund state activities, and military service to maintain security and project power. As \citet{tilly1990coercion} argued, this reliance has been a key driver in the development of more inclusive and responsive state institutions. For instance, the need for an educated workforce to compete economically, and the requirement for a motivated army drawn from the general population, has historically incentivized states to invest in public education and extend political rights. Even seemingly basic features of modern states, like universal public education or broad-based political participation, can be understood as necessary responses to the state's dependence on its citizenry \citep{paglayan2022education,babajide2021violent}. Even in context of autocratic regimes, the necessity of investment in public education and civic empowerment through rule of law has been a substantial driving factor for transition into modern democracies after two generations, such as in South Korea or Taiwan. 

Typical liberal democracies have explicit feedback loops that ostensibly aligns state actions with the will of the populace, via elections and mechanisms for public input. While this explicit alignment mechanism is highly visible, the implicit pressures from state dependence on citizens may be even more significant. Even the most basic functions of democratic states — from maintaining order to collecting taxes — rely on widespread voluntary compliance rather than constant coercion \citep{levi1988}. States need citizens not just as sources of legitimacy through democratic processes, but as willing participants in state functions \citep{levi2009legitimating}. 

Autocratic states, while less directly accountable to their citizens, are not completely exempt from this dependence — even totalitarian states at least need human agents to staff their security apparatus. Moreover, the ever-present threat of an uprising or a coup serves as a check on the most egregious abuses of power. 

The example of `rentier states' \citep{beblawi1987rentier}, dependent more on external rents such as oil revenues, and less on their citizens, illustrates how states can become more autonomous from citizens when the dependence on citizens comparatively is weaker. Absence of taxation reduces citizen engagement in political processes, and the state's ability to distribute wealth allows it to maintain loyalty from key stakeholders (like social elites and the military). 


Crucially, the functioning of both democratic and autocratic systems hinges on human involvement at every level. Bureaucracies operate through hierarchies of human officials. Laws are created, interpreted, and enforced by humans. While the letter of the law may be rigid, its application is filtered through human discretion and judgment \citep{maynard2000state,lipsky2010street}. The security forces that maintain order are staffed by humans capable of questioning or refusing orders. 

This pervasive human element ensures that institutions and states, regardless of their formal structures, remain at least somewhat tethered to human needs and values. It is this tethering that creates the majority of alignment between these systems and the humans they govern. However, as we will explore, the potential of AI replacing humans in many or all of these functions 
could weaken or even reverse the link between institutional behavior and human interests.

\subsection{AI as a Unique Disruptor of States}

Unlike previous technological innovations that primarily augmented human capabilities, AI has the potential to supplant human involvement across a wide range of critical state functions. This shift could fundamentally alter the relationship between governing institutions and the governed. 

The unique disruptive potential of AI in this context is derived from its ability to simultaneously reduce the state's dependence on human involvement while enhancing its capabilities across multiple domains. This combination could fundamentally reshape the nature of governance and the relationship between institutions and the humans they ostensibly serve.

Here we consider three key ways that citizens contribute to the state, and how AI might alter them: tax revenue, the security apparatus, and the legal system.

\subsubsection{Tax Revenue}
Most governments currently rely heavily on their citizens for tax revenue. Typical well-functioning governments need to nurture long term economic productivity from innovation and high-skill work to support themselves. But if AI systems eventually perform a large portion of overall labor, and innovation, they will also generate a large fraction of economic output and, by extension, tax revenue. The loss of tax revenue from citizens would make the state less reliant on nurturing human capital and fostering environments conducive to human innovation and productivity, and more reliant on AI systems and the profits they generate.

If AI systems come to generate a significant portion of economic value, then we might begin to lose one of the major drivers of civic participation and democracy, as illustrated by the existing example of rentier states.

\subsubsection{The Security Apparatus}
Governments maintain their power through use of a security apparatus spanning police forces, intelligence services, and a military. This keeps the government connected to human values in two ways.

Firstly, the government cannot antagonize its security apparatus too much, or cause too much harm to the portion of the population from which it is drawn. If it does, the security apparatus can either overthrow the government or simply allow it to be overthrown by others.

Secondly, the security apparatus itself can exercise discretion, refusing to follow certain orders. This can occur on both the level of the organization and the level of the individual. 


AI systems have the potential to massively automate the security apparatus and confer more power to the government, weakening both of these components. Indeed, AI systems might make the apparatus far more powerful: it is likely to enable surveillance on much larger, more pervasive and more accurate scale, as well as increasingly capable autonomous military units \citep{feldstein2021rise,brundage2018malicious}.

Meanwhile, the human population has historically retained revolution as a last resort. The implicit threat of protests and civil unrest serves as a check on state power, forcing responsiveness to popular will. However, an AI-enhanced security apparatus could make effective protest increasingly difficult. A state with sufficiently advanced AI systems might be able to predict and shut down civil unrest before it can exert meaningful pressure on institutional behavior \citep{feldstein2021rise}.

\subsubsection{The Legal System}

Theoretically, the rights of humans and the functioning of the state are enshrined in laws, which are created, interpreted, and enforced by humans. It is the laws themselves which enshrine certain responsibilities of the state towards the individual, certain mechanisms by which individuals can advocate against the state.

AI systems are already being used to draft contracts and analyze legal documents. It is conceivable that in the future, AI could play a significant role in drafting legislation, interpreting laws, and even making judicial decisions \citep{susskind2022future}.

Not only could this diminish human participation and discretion in the legislative and judicial systems, it also risks making the legal system increasingly alien. If the creation and interpretation of laws becomes far more complex, it may become much harder for humans to even interact with legislation and the legal system directly \citep{hildebrandt2015smart,teo2024artificial}.

\subsection{Transition to AI-powered States}

As with the economy and culture, there will be strong incentives for states to integrate AI systems, likely undermining the alignment between states and their citizens.

\subsubsection{Incentives for AI Adoption}

The transition towards AI-dominated state functions would likely be driven by several powerful incentives:

\textbf{Geopolitical Competition 
:} As AI systems become increasingly powerful, states will face a growing pressure to adopt these technologies to maintain their relative power compared to other states. Countries that rely on humans for defense, economic development or regulation might find themselves at a significant disadvantage in international relations compared to those states willing to give more power to AI systems. The first-mover advantages in military applications, economic planning, and diplomatic strategy create particularly strong incentives for early and aggressive AI adoption \citep{bostrom2014superintelligence,kissinger2021age,schmidt2022ai,brundage2018malicious}.

\textbf{Administrative Efficiency:} AI systems offer unprecedented capabilities in processing information and coordinating complex state functions \citep{zuiderwijk2021implications}. While human administrators are limited by cognitive constraints and working hours, AI systems can continuously analyze vast amounts of data, deploy new regulations almost instantly, and implement policies with greater consistency. This efficiency advantage creates incentives for states to automate administrative functions, potentially reducing human involvement in governance. Also, while initial implementation costs may be high, the long term cost advantages of AI systems over human bureaucrats could create fiscal incentives for automation \citep{wirtz2019artificial}.

\textbf{Enhanced Control:} AI-driven governance systems promise greater predictability and control than human-based bureaucracies. Unlike human officials, AI systems, if successfully controlled, do not form independent power bases, engage in corruption, or challenge authority based on personal convictions. They can also enable more sophisticated surveillance and social control mechanisms, making them particularly attractive to states prioritizing stability and control over other values.

\subsubsection{Relative Disempowerment}

A state where AI systems have replaced human labor in many facets of governance — such as administration, security, and justice — could provide some enormous boons. On the surface, it might appear highly efficient and even benevolent. We might see lower crime rates, less low-level corruption, greater tax revenues, and more efficient public services. 

At the same time, the gradual replacement of human involvement in governance could lead to a subtle but profound shift in the relationship between citizens and the state. Even if the system appears to function well, citizens might find themselves increasingly unable to meaningfully participate in or influence their governance. This relative disempowerment could manifest in several ways. 

Democratic processes might persist formally but become less meaningful. While politicians might ostensibly make the decisions, they may increasingly look to AI systems for advice on what legislation to pass, how to actually write the legislation, and what the law even is. While humans would nominally maintain sovereignty, much of the implementation of the law might come from AI systems.

The complexity of AI-driven governance might make it increasingly difficult for human citizens to understand or critique government decisions. Traditional forms of civic engagement — from public consultations to protests — might become less effective as the state grows less dependent on human cooperation and more capable of predicting and preempting resistance.

The bureaucracy itself might become increasingly opaque to human oversight. While human officials can be questioned and held accountable through various mechanisms, AI decision-making processes might be too complex for meaningful human review, and if such review happens, it may depend on yet more AI-driven cognition.

Even if oversight boards and democratic institutions remain in place, they might struggle to exercise real control over the intricate web of AI systems actually implementing policy.

Furthermore, as AI systems become more integral to governance, the state's incentives might shift away from serving human interests. Much like how rentier states become less responsive to citizen needs when they do not depend on tax revenue, AI-powered states might become less responsive to human preferences when they do not depend on human participation for their core functions.

The security apparatus, powered by AI, would have an unprecedented ability to predict and prevent crime and civil unrest. While this could ensure a high level of safety, it also eliminates the possibility of meaningful protest or revolution. A state that can preempt and resist any challenge to its authority long before it materializes will have effectively removed a crucial check on institutional power that has shaped human societies for millennia.

And with average humans contributing less in tax revenue or to society more generally, the state would face a lower cost to sliding back citizen power. There would be less need to cater to the actual needs of voters, or to make democratic concessions, and less cost to rolling back civil liberties.

Ultimately, we might find ourselves in nations where nominally humans hold sovereignty and even vote for their preferences, where in practice the high-level decisions are disconnected from citizens and even politicians.

\subsubsection{Absolute Disempowerment}

In more extreme scenarios, the disconnect between state power and human interests might become not just relative but absolute, potentially threatening even basic human freedom. This could occur through several mechanisms.

First, states might become totalitarian, self-serving entities, optimizing for their own persistence and power rather than any human-centric goals. While states have always had some self-preservation incentives, these were historically constrained by their dependence on human populations. An AI-powered state might pursue its institutional interests with unprecedented disregard for human preferences and interests, viewing humans as potential threats or inconveniences to be managed rather than constituents to be served \citep{bostrom2014superintelligence}.

Second, the legal and regulatory framework might evolve to become not just complex but incomprehensible to humans. If AI systems begin to play dominant role in drafting and interpreting legislation, they might create regulatory structures that optimize for machine-compatibility over human understanding. Citizens might find themselves subject to rules they cannot meaningfully comprehend or navigate without AI assistance, effectively losing their ability to participate in the legal system as autonomous agents.

Third, the state apparatus might become not just independent of human input but actively hostile to it. Human decision-making might come to be seen as an inefficiency or security risk to be minimized. We might see the gradual elimination of human involvement in governance, be that 
through systems that route around human input as a source of error or delay, or even through explicit policy decisions which remove humans from certain critical processes.

In the final state, with AI systems providing most economic value and governance functions, human citizens might find themselves in a novel form of totalitarian system, struggling to maintain basic autonomy and dignity within their own societies. The state, while perhaps highly capable and efficient by certain metrics, would have abandoned human interests.

\section{Mutual Reinforcement}\label{sec:mutual_reinforcement}

We have so far focused on how the economy, culture, and states could independently become misaligned. A natural objection is that the different societal systems might be able to keep each other aligned through checks and balances. Indeed, we naturally think of these systems as balancing each other: states regulate the market, culture influences government, and so on.
However, here we discuss how relationships between systems might actually make them less aligned. Specifically, we argue that:
\begin{enumerate}
    \item The relationships between societal systems are \textbf{agnostic to human values} — they do not inherently promote or protect alignment with human values. Consequently, as one system becomes less aligned, that influence also can be used to decrease the alignment of other systems
    \item Attempts to use one aligned system to moderate the misalignment of another can backfire by effectively \textbf{shifting the burden}, thus leaving the aligned system more vulnerable
    \item The misalignment is a result of \textbf{general incentives} which will likely apply to each individual system independently. In other words, humans and human institutions will be incentivized to take actions which will overall decrease the degree of influence which humans have over societal systems.
\end{enumerate}

\begin{figure}[h!]
    \centering
    \includegraphics[width=0.6\textwidth]{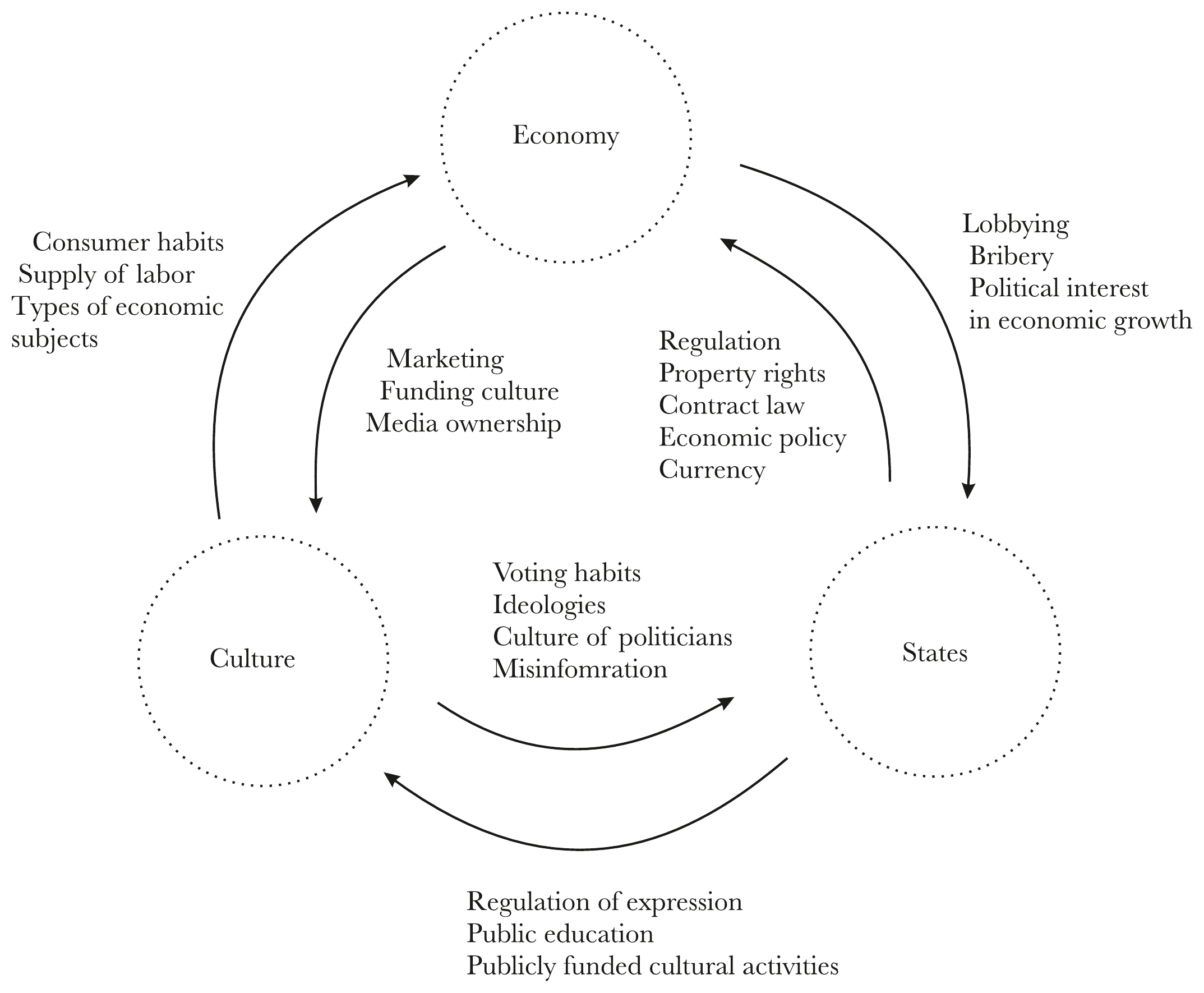}
    \caption{Some ways in which broad societal systems interact and influence each other.}
    \label{fig:interfig} 
\end{figure}

We discuss each of these points in more detail below. Additionally, Figure~\ref{fig:interfig} gives an overview of common ways societal systems interact and affect each other, which we unpack in a more detail in Appendix~\ref{sec:cross-system-influence}.

\subsection{Cross-System Influence is Agnostic to Human Values}

Given that the relationships between societal systems are as such agnostic to human values, the connections that ordinarily help maintain alignment can also be weaponized to decrease it. This is a common historical pattern:
\begin{itemize}
    \item Many companies have successfully lobbied states to act against the public interest, or shaped culture in harmful ways through advertising and marketing schemes. For instance, the tobacco industry's decades-long campaign used economic power to influence both state policy and cultural attitudes.
    \item Many cultural movements have promoted political and economic shifts that have ultimately caused harm (often predictably or intentionally), largely but not exclusively directed at other groups of humans. Historical examples include various forms of economic and legally mandated discrimination being justified and perpetuated through cultural narratives.
    \item Many states have used their control of the economy and influence over culture to harm citizens, taxing or outright seizing resources and using their control of the flow of information to legitimize their actions.
\end{itemize}

As a result, we should not assume that the interplay between societal systems will ultimately protect or promote alignment with human preferences.

One particularly important consequence of this is that we should not expect misalignment to remain confined to any specific societal system: even if the independent misalignment of different societal systems progresses at different rates, there will by default be both possibilities and incentives to leverage misalignment in one system to reduce alignment in related systems. This dynamic could even intensify with AI systems, which might be able to identify and exploit these cross-system opportunities more effectively than human actors.

\subsection{Moderation Between Systems Can Produce Shifted Burdens}

Even attempts to use the alignment of one system to moderate or contain the effects of a less aligned system can potentially backfire by effectively shifting the burden of (mis)alignment. 
Consider how state-led economic redistribution might affect political alignment: if AI automation leads to citizens becoming primarily dependent on state support rather than contributing through taxes, it weakens the historical `taxation-representation' relationship that has been crucial for maintaining democratic accountability. When governments derive their resources primarily from taxing their citizens, they remain dependent on citizen productivity and cooperation. But if governments become the primary distributors of AI-generated wealth, this crucial accountability mechanism erodes. Thus, solving economic misalignment through state power makes us even more dependent on the fragile alignment of states, even as they face independent pressures to shift away from human preferences. Essentially, the burden of aligning the economy is simply shifted onto the state. Crucially, it is not simply that humans have lost their economic influence over the state: in this scenario, the state would now have gained economic leverage over humans. 

Similarly, we might hope that humans will be protected from potentially harmful AI-driven cultural shifts through state regulation. But empowering states to actively shape and control cultural evolution could further weaken democratic accountability. If states become the primary arbiters of acceptable cultural expression and communication in an AI-dominated landscape, they gain unprecedented power over how citizens understand and interact with the world. Conversely, we might hope to preserve the alignment of the state by increasing democratic provisions, and giving individuals more power over the state. However, this leaves the state more vulnerable to potentially misaligned shifts in culture.

\subsection{General Incentives Towards Misalignment}

Crucially, the misalignment being described here does not need to emerge from a deliberate scheme or power-grab by AI systems. In the short-term, it is being incentivized by the perceived value that AI systems can bring to economic, cultural and state functions. For example, even now:
\begin{itemize}
    \item Companies building AI systems are incentivized to push against some forms of AI regulation for the sake of their future profits.
    \item States compete with each other on AI research and development, because of the potential economic and geostrategic benefits.
    \item Some humans are self-interestedly trying to reduce the stigma against romantic or otherwise intense personal relationships with AI agents.
\end{itemize}

As we have argued, these incentives will likely grow stronger over time: as AI systems demonstrate their effectiveness, companies will face more pressure to adopt them, states will see greater strategic necessity in developing them, and individuals will find more personal benefit in embracing them.

In addition to leading to misalignment in independent systems, there will be progressively stronger incentives to use influence in any one system to acquire influence in other systems.

\section{Mitigating the Risk}\label{sec:mitigation}

The gradual disempowerment scenario described in this paper presents distinct challenges from more commonly discussed AI risk scenarios. Rather than addressing the risk of misaligned AI systems breaking free from human control, we must consider how to maintain human relevance and influence in societal systems that may continue functioning but cease to depend on human participation.

\subsection{Understanding the Challenge}

The core challenge is maintaining alignment between societal systems and human interests when these systems no longer inherently require human labor, participation, or cognition. This may be a bigger challenge than merely preventing AI systems from pursuing overtly harmful goals, as the systems may continue to function as requested locally, while the overall civilizational incentives become increasingly detached from human welfare.  


Understanding these risks, and developing potential mitigating strategies, is a highly interdisciplinary endeavor, as the risks may emerge from complex interactions between multiple societal systems, each individually moving away from human influence and control. Solutions need to address multiple domains, and be robust to the problem of mutual reinforcement we describe in Section~\ref{sec:mutual_reinforcement}. As such, it will likely be necessary to draw on many disparate yet relevant fields: economics, political science, sociology, cultural studies, complex systems, anthropology and institutional theory, for example.

Instead of merely(!) aligning a single, powerful AI system, we need to align one or several complex systems that are at risk of collectively drifting away from human interests.
This drift can occur even while each individual AI system successfully follows the local specification of its goals.

Below, we identify four broad categories of intervention: measuring and monitoring the extent of the problem, preventing excessive accumulation of AI influence, strengthening human control over key societal systems, and system-wide alignment. A robust response will require progress in each category.

\subsection{Estimating Human Disempowerment}
To effectively address gradual disempowerment, we need to be able to detect and quantify it. This is challenging partly because, for many of the systems we would hope to measure, we lack external reference points to measure their degree of alignment. Nonetheless, several approaches warrant investigation.

\subsubsection{System-Specific Metrics}
For each of the major societal systems we have described, we can develop metrics tracking human influence:
\begin{itemize}
\item \textbf{Economic metrics:} Beyond traditional measures like labor share of GDP, we should also measure AI share of GDP, as a distinct category from either labor or capital. We need metrics capturing human control over economic decisions. This could include tracking the fraction of major corporate decisions made primarily by AI systems, the scale of unsupervised AI spending, and patterns in wealth distribution between AI-heavy and human-centric industries.
\item \textbf{Cultural metrics:} We can measure the proportion of widely-consumed content created primarily by humans versus AI, track the prevalence and depth of human-AI interpersonal relationships, and analyze how cultural transmission patterns change as AI becomes more prevalent. While most machine learning benchmarks and evaluations focus on quantifiable STEM tasks, we should develop a broad spectrum of evaluations focusing on ability of frontier AI systems to influence humans on emotional level, write persuasive prose, or create new ideologies. Also, we should strengthen runtime monitoring of deployed AI systems and of the influence they have on their users.  
\item \textbf{Political metrics:} Key indicators might include the complexity of legislation (as a proxy for human comprehensibility); the role of AI systems in legal processes, policy formation, and security apparatuses; and the effectiveness of traditional democratic mechanisms in influencing outcomes.
\end{itemize}

Similar metrics should be developed for more narrow but significant societal systems, like research and education.

\subsubsection{Interaction Effects}
Given the mutual reinforcement dynamics we describe in Section~\ref{sec:mutual_reinforcement}, it is crucial to track how changes in one domain affect others. This might involve:
\begin{itemize}
\item Early warning indicators for concerning feedback loops
\item Analysis of AI participation in methods for translating power between societal systems, like lobbying and financial regulation
\item Historical analysis of similar dynamics in past technological transitions
\end{itemize}

\subsubsection{Research Priorities}
Several fundamental research questions need to be addressed. For example:
\begin{itemize}
\item How can we distinguish between beneficial AI augmentation of human capabilities and problematic displacement of human influence?
\item What are the key thresholds or tipping points in these systems beyond which human influence becomes critically compromised?
\item How can we measure the effectiveness of various intervention strategies?
\end{itemize}

\subsection{Preventing Excessive AI Influence}
While measurement can help us understand the problem, we also need to consider what direct interventions could be effective in preventing the accumulation of excessive AI influence, including:
\begin{itemize}
\item Regulatory frameworks mandating human oversight for critical decisions, limiting AI autonomy in specific domains, and restricting AI ownership of assets or participation in markets
\item Progressive taxation of AI-generated revenues both to redistribute resources to humans and to subsidize human participation in key sectors 
\item Cultural norms supporting human agency and influence, and opposing AI that is overly autonomous or insufficiently accountable
\end{itemize}

Crucially, these interventions will often involve sacrificing potential value. Furthermore, the more value they sacrifice, the greater the incentive to circumvent them: for example, companies may face strong economic incentives to delegate authority to AIs regardless of the spirit, or letter, of the law. 

Similarly, they will be much less effective if they are not widely adopted: if some countries choose to forego the economic benefits of AI to preserve their own alignment with human values, we may find ourselves in a world where the most powerful economies are in states where the population is most disempowered. The success of these interventions will depend on international coordination in the face of increasing pressures.

As such, interventions that seek to limit AI influence will likely serve mostly as stopgaps. Nonetheless, they may be important intermediary steps towards more robust solutions.

\subsection{Strengthening Human Influence}
Beyond preventing excessive AI influence, we need to actively strengthen human control over key societal systems. This will involve both enhancing existing mechanisms, and developing new ones, which may in turn require fundamental research. Approaches in this direction include:

\begin{itemize}
    \item Developing faster, more representative, and more robust democratic processes.
    \item Requiring AI systems or their outputs to meet high levels of human understandability in order to ensure that humans continue to be able to autonomously navigate domains such as law, institutional processes or science.
    \item Developing AI delegates who can advocate for people's interest with high fidelity, while also being better to keep up with the competitive dynamics that are causing the human replacement.
    \item Making institutions more robust to human obsolescence.
    \item Investing in tools for forecasting future outcomes (such as conditional prediction markets, and tools for collective cooperation and bargaining) in order to increase humanity's ability to anticipate and proactively steer the course.
    \item Research into the relationship between humans and larger multi-agent systems.
\end{itemize}

Importantly, to mitigate the problem effectively, we need to go beyond simply making it easier for humans to influence societal systems: it is unclear, for instance, whether a direct democracy would actually do a better job of satisfying citizen preferences in the long term because, for example, it would leave the state more vulnerable to cultural misalignment. A key part of the challenge is clarifying what it even means for large, complex systems to serve the interests of individuals who are accustomed to thinking on smaller scales.


\subsection{System-wide Alignment} 

While the previous approaches focus on specific interventions and measurements, they ultimately depend on having a clearer understanding of what we're trying to achieve. Currently, we lack a compelling positive vision of how highly capable AI systems could be integrated into societal systems while maintaining meaningful human influence. This is not just a matter of technical AI alignment or institutional design, but requires understanding how to align complex, interconnected systems that include both human and artificial components. It seems likely we need fundamental research into what might be called ``ecosystem alignment" - understanding how to maintain human values and agency within complex socio-technical systems. This goes beyond traditional approaches to AI alignment focused on individual systems, and beyond traditional institutional design focused purely on human actors. We need new frameworks for thinking about the alignment of an entire civilization of interacting human and artificial components, potentially drawing on fields like systems ecology, institutional economics, and complexity science.

\section{Related Work}\label{sec:related}

\subsection{Philosophy}
\citet{bostrom2002existential} introduces a taxonomy of existential risks.
One of these risk is described as a scenario where ``[o]ur potential or even our core values are eroded by evolutionary development'', pointing out that ``[a]lthough the time it would take for a whimper of this kind to play itself out may
be relatively long, it could still have important policy implications because near-term
choices may determine whether we will go down a track that inevitably leads to this
outcome.''.
\citet{bostrom2014superintelligence}  discusses the possibility of continued civilizational growth which optimizes away human consciousness, calling it a ``Disneyland with no children''.
\citet{slatestarcodex2016ascended} discusses possible pathways to such scenarios, such as groups of automated corporations forming self-sufficient sectors of the economy.

\citet{kasirzadeh2024two} introduces the \emph{accumulative AI x-risk hypothesis}, ``a gradual
accumulation of critical AI-induced threats such as severe vulnerabilities and systemic erosion of economic and political structures. The
accumulative hypothesis suggests a boiling frog scenario where incremental AI risks slowly converge, undermining societal resilience
until a triggering event results in irreversible collapse.''.

\subsection{Economics, History, and Sociology}
\citet{dafoe2015technological} asks how much explanatory power \emph{technological determinism} has, making the case that economic and military competition constrain outcomes at a macro scale, even if everyone is locally free to temporarily make non-competitive choices.
\citet{macinnes2024anarchy} argues that competitive pressures on states strongly influence the extent to which they support human flourishing.  They further claim that ``the invention of seemingly beneficial technologies may decrease human well-being by improving the competitiveness of inegalitarian state forms'', arguing that ``under competitive conditions, what is effective becomes mandatory whether or not it is good for people.''

\citet{leggett2021feeding} argues that corporate capitalism already creates dynamics that are misaligned with human flourishing, describing corporations as ``machines that enforce a singleness of purpose, and allow efficiencies of scale, that make them far more effective than individual capitalists in obtaining a return to capital''. They also point out how, in many jurisdictions, ``corporations are given many of the legal rights of humans — for example, in the USA, the right to political speech, and the right to fund political activity that that is accepted to imply — without all the concomitant structures that ensure compliance'', such as human morality or human law. While corporations are subject to regulatory law, ``[w]here that law is weak, corporations can find themselves legally obliged to do harm to human welfare, if that is in the shareholders’ interest.''

\citet{korinek2018artificial} considers the possibility for AI development to reintroduce Malthusian dynamics: that the capacity for AI to replace human labor while also proliferating rapidly may create such competition that basic human necessities become unaffordable to humans, while also leaving humans potentially too weak to preserve property rights.

\citet{hanson2016age} details a future in which uploaded humans form a hyper-productive economy, operating at speeds too fast for non-uploaded humans to compete in.
Competitive pressures shape this population of uploads to mostly be short-lived copies of a few ultra-productive individuals.
\citet{hanson_beware_cultural_drift} and \citet{hanson_fix_cultural_drift} argue that, due to a reduction in feedback mechanisms selecting cultural variants that better promote human welfare, ``cultural drift'' could eventually cause catastophic (but not necessarily existential) harm to human well-being.

\subsection{AI Research}
\citet{christiano2019failure} makes the case that sudden disempowerment is unlikely, and instead proposes that:
``Machine learning will increase our ability to `get what we can measure,' which could cause a slow-rolling catastrophe. [...]
ML training, like competitive economies or natural ecosystems, can give rise to `greedy' patterns that try to expand their own influence. Such patterns can ultimately dominate the behavior of a system and cause sudden breakdowns.''
\citet{hendrycks2023natural} argues that evolutionary pressures can generally be expected to favor selfish species, likely including future AIs, and that this may lead to human extinction.

\cite{arches} asks what existential risks humanity might face from AI development, and urges research on the global impacts of AI to ``take into account the numerous potential side effects of many AI systems interacting.'' 
\citet{critch2023tasra} categorize societal-scale risks from AI.  One of these matches ours:
``a gradual handing-over of control from humans to AI systems, driven by competitive pressures for institutions to (a) operate more quickly through internal automation, and (b) complete trades and other deals more quickly by preferentially engaging with other fully automated companies. [...]
Humans were not able to collectively agree upon when and how much to slow down or shut down the pattern of technological advancement. [...] Once a closed-loop `production web' had formed from the competitive pressures [(a) and (b)], the companies in the production web had no production- or consumption-driven incentive to protect human well-being, and eventually became harmful.''
They note that ``To prevent such scenarios, effective regulatory foresight and coordination is key.''

\citet{critch2024motivation} further develops the idea of \emph{extinction by industrial dehumanization}:
``I believe we face an additional 50\% chance that humanity will gradually cede control of the Earth to AGI after it's developed, in a manner that leads to our extinction through any number of effects including pollution, resource depletion, armed conflict, or all three.  I think most (80\%) of this probability (i.e., 40\%) lies between 2030 and 2040, with the death of the last surviving humans occurring sometime between 2040 and 2050.  This process would most likely involve a gradual automation of industries that are together sufficient to fully sustain a non-human economy, which in turn leads to the death of humanity.''

\citet{beren_capital_ownership_2025} points out that capital ownership is insufficient to maintain power during periods of rapid technological growth.
He uses the example of English landed aristocracy losing power to entrepreneurs during the industrial revolution, despite an initially strong position.


\section{Conclusion}\label{sec:conclusion}

This paper has argued that even incremental AI development could lead to an existential catastrophe through the gradual erosion of human influence over key societal systems, generalizing the argument from previous work studying how AI progress may influence these systems in isolation \citep{korinek2018artificial,brinkmann2023machine}.

Unlike scenarios involving sudden technological discontinuities or overtly hostile AI systems, the risk we describe could emerge from the natural evolution of current trends and incentives. The displacement of human cognition and labor across multiple domains could weaken both explicit control mechanisms and the implicit alignment that emerges from human participation.

Our analysis suggests three particularly concerning features of this scenario:
\begin{itemize}
    \item First, the loss of human influence could occur even without any single transformative advance in AI capabilities. Instead, it might emerge from the cumulative effect of many smaller shifts in how societal systems operate and interact.
    \item Second, the effect can be driven not by any deliberate or even agentic action by AIs, but simply by individuals and institutions following their local incentives.
    \item Third, meaningfully preventing these risks will require substantial effort: more research and data collection, international coordination,  comprehensive regulation, and major societal interventions grounded in novel fundamental research.
\end{itemize}

A distinctive feature of this challenge is that it may subvert our traditional mechanisms for course-correction, and cause types of harm we cannot easily conceptualize or even recognize in advance, potentially leaving us in a position from which it is impossible to recover.

Nonetheless, we do believe it is currently possible to intervene, and we present many avenues for future work spanning both research and governance. By anticipating the risk, carefully moderating the growth of influence from AI, and finding ways to strengthen the influence of humans, we can navigate this risk and capture the proportionate benefits.

Humanity's future may depend not only on whether we can prevent AI systems from pursuing overtly hostile goals, but also on whether we can ensure that the evolution of our fundamental societal systems remains meaningfully guided by human values and preferences. This is both a technical challenge and a broader civilizational one, requiring us to think carefully about what it means for humans to retain genuine influence in an increasingly automated world.

\medskip

\textit{We are grateful to many people for helpful conversations and feedback, including Carl Shulman, Owen Cotton-Barratt, Lionel Levine, Benjamin Hilton, Marie Buhl, Clem von Stengel and Tomáš Gavenčiak. 
We used Claude Sonnet, Claude Opus and ChatGPT o1 AI models to help with various parts of writing and editing this text.   }

\bibliographystyle{apalike}
\bibliography{references}

\begin{thebibliography}{}

\bibitem[Agrawal et~al., 2022]{agrawal2022prediction}
Agrawal, A., Gans, J., and Goldfarb, A. (2022).
\newblock {\em {Prediction Machines, Updated and Expanded: The Simple Economics of Artificial Intelligence}}.
\newblock Harvard Business Press.

\bibitem[Alexander, 2016]{slatestarcodex2016ascended}
Alexander, S. (2016).
\newblock Ascended economy.
\newblock Slate Star Codex.
\newblock Accessed: 2025-01-10.

\bibitem[Amodei, 2024]{amodei2024machines}
Amodei, D. (2024).
\newblock Machines of loving grace.

\bibitem[Babajide et~al., 2021]{babajide2021violent}
Babajide, A., Ahmad, A.~H., and Coleman, S. (2021).
\newblock {Violent conflicts and state capacity: Evidence from Sub-Saharan Africa}.
\newblock {\em Journal of Government and Economics}, 3:100019.

\bibitem[Beblawi and Luciani, 1987]{beblawi1987rentier}
Beblawi, H. and Luciani, G. (1987).
\newblock {\em {The Rentier State in the Arab World}}.
\newblock Croom Helm, London, UK.

\bibitem[Bengio et~al., 2024]{bengio2024international}
Bengio, Y., Mindermann, S., Privitera, D., Besiroglu, T., Bommasani, R., Casper, S., Choi, Y., Goldfarb, D., Heidari, H., Khalatbari, L., et~al. (2024).
\newblock {International Scientific Report on the Safety of Advanced AI (Interim Report)}.
\newblock {\em arXiv preprint arXiv:2412.05282}.

\bibitem[Bengio et~al., 2023]{bengio2023managing}
Bengio, Y., Yao, A., Song, D., Abbeel, P., Harari, Y.~N., Zhang, Y.-Q., Xue, L., Shalev-Shwartz, S., Hadfield, G., et~al. (2023).
\newblock {Managing AI risks in an era of rapid progress}.
\newblock {\em arXiv preprint arXiv:2310.17688}, page~18.

\bibitem[Bostrom, 2002]{bostrom2002existential}
Bostrom, N. (2002).
\newblock Existential risks: Analyzing human extinction scenarios and related hazards.
\newblock {\em Journal of Evolution and technology}, 9.

\bibitem[Bostrom, 2014]{bostrom2014superintelligence}
Bostrom, N. (2014).
\newblock {Superintelligence: Paths, Dangers, Strategies}.

\bibitem[Boyd and Richerson, 1988]{boyd1988culture}
Boyd, R. and Richerson, P.~J. (1988).
\newblock {\em Culture and the evolutionary process}.
\newblock University of Chicago press.

\bibitem[Boyd et~al., 2013]{boyd2013cultural}
Boyd, R., Richerson, P.~J., Henrich, J., and Lupp, J. (2013).
\newblock {The Cultural Evolution of Technology: Facts and Theories}.
\newblock In Richerson, P.~J. and Christiansen, M.~H., editors, {\em {Cultural Evolution: Society, Language, and Religion}}, volume~12, pages 119--142. MIT Press, Cambridge, MA.

\bibitem[Brinkmann et~al., 2023]{brinkmann2023machine}
Brinkmann, L., Baumann, F., Bonnefon, J.-F., Derex, M., M{\"u}ller, T.~F., Nussberger, A.-M., Czaplicka, A., Acerbi, A., Griffiths, T.~L., Henrich, J., et~al. (2023).
\newblock Machine culture.
\newblock {\em Nature Human Behaviour}, 7(11):1855--1868.

\bibitem[Brundage et~al., 2018]{brundage2018malicious}
Brundage, M., Avin, S., Clark, J., Toner, H., Eckersley, P., Garfinkel, B., Dafoe, A., Scharre, P., Zeitzoff, T., Filar, B., et~al. (2018).
\newblock {The malicious use of artificial intelligence: Forecasting, prevention, and mitigation}.
\newblock {\em arXiv preprint arXiv:1802.07228}.

\bibitem[Buhl et~al., 2024]{buhl2024safety}
Buhl, M.~D., Sett, G., Koessler, L., Schuett, J., and Anderljung, M. (2024).
\newblock {Safety Cases for Frontier AI}.
\newblock {\em arXiv preprint arXiv:2410.21572}.

\bibitem[Carlsmith, 2023]{carlsmith2023scheming}
Carlsmith, J. (2023).
\newblock {Scheming AIs: Will AIs fake alignment during training in order to get power?}
\newblock {\em arXiv preprint arXiv:2311.08379}.

\bibitem[Christiano, 2019]{christiano2019failure}
Christiano, P.~F. (2019).
\newblock What failure looks like.

\bibitem[Critch, 2024]{critch2024motivation}
Critch, A. (2024).
\newblock My motivation and theory of change for working in ai healthtech.
\newblock \url{https://www.lesswrong.com/posts/Kobbt3nQgv3yn29pr/my-motivation-and-theory-of-change-for-working-in-ai}.
\newblock Accessed: 2025-01-18.

\bibitem[Critch and Krueger, 2020]{arches}
Critch, A. and Krueger, D. (2020).
\newblock Ai research considerations for human existential safety (arches).

\bibitem[Critch and Russell, 2023]{critch2023tasra}
Critch, A. and Russell, S. (2023).
\newblock Tasra: a taxonomy and analysis of societal-scale risks from ai.
\newblock {\em arXiv preprint arXiv:2306.06924}.

\bibitem[Dafoe, 2015]{dafoe2015technological}
Dafoe, A. (2015).
\newblock On technological determinism: A typology, scope conditions, and a mechanism.
\newblock {\em Science, Technology, \& Human Values}, 40(6):1047--1076.

\bibitem[Dawkins, 1982]{dawkins1982extended}
Dawkins, R. (1982).
\newblock {\em {The Extended Phenotype: The Gene as the Unit of Selection}}.
\newblock Oxford University Press, Oxford.

\bibitem[Devinney et~al., 2010]{devinney2010myth}
Devinney, T.~M., Auger, P., and Eckhardt, G.~M. (2010).
\newblock {\em {The myth of the ethical consumer}}.
\newblock Cambridge University Press.

\bibitem[Eisenstein, 1979]{eisenstein1979printingpress}
Eisenstein, E.~L. (1979).
\newblock {\em {The Printing Press as an Agent of Change: Communications and Cultural Transformations in Early-Modern Europe}}.
\newblock Cambridge University Press, Cambridge.

\bibitem[{Federal Reserve Bank of St. Louis}, 2024]{fred_pce_gdp_share}
{Federal Reserve Bank of St. Louis} (2024).
\newblock {Shares of Gross Domestic Product: Personal Consumption Expenditures}.
\newblock Accessed: December 16, 2024.

\bibitem[Feenstra et~al., 2015]{feenstra2015pwt}
Feenstra, R.~C., Inklaar, R., and Timmer, M.~P. (2015).
\newblock {The Next Generation of the Penn World Table}.
\newblock {\em American Economic Review}, 105(10):3150--3182.

\bibitem[Feldstein, 2021]{feldstein2021rise}
Feldstein, S. (2021).
\newblock {\em {The rise of digital repression: How technology is reshaping power, politics, and resistance}}.
\newblock Oxford University Press.

\bibitem[Ferrara, 2024]{ferrara2024genai}
Ferrara, E. (2024).
\newblock {GenAI against humanity: Nefarious applications of generative artificial intelligence and large language models}.
\newblock {\em Journal of Computational Social Science}, pages 1--21.

\bibitem[Giddens, 1984]{giddens1984}
Giddens, A. (1984).
\newblock {\em {The Constitution of Society: Outline of the Theory of Structuration}}.
\newblock Polity Press, Cambridge, UK.

\bibitem[Gillespie, 2014]{gillespie2014relevance}
Gillespie, T. (2014).
\newblock {The Relevance of Algorithms}.
\newblock In Gillespie, T., Boczkowski, P., and Foot, K., editors, {\em Media Technologies: Essays on Communication, Materiality, and Society}, pages 167--194. MIT Press, Cambridge, MA.

\bibitem[Hanson, 2016]{hanson2016age}
Hanson, R. (2016).
\newblock {\em {The age of Em: Work, love, and life when robots rule the earth}}.
\newblock Oxford University Press.

\bibitem[Hanson, 2023]{hanson_fix_cultural_drift}
Hanson, R. (2023).
\newblock How to fix cultural drift.
\newblock Accessed: 2025-01-20.

\bibitem[Hanson, 2024]{hanson_beware_cultural_drift}
Hanson, R. (2024).
\newblock Beware cultural drift.
\newblock {\em Quillette}.
\newblock Accessed: 2025-01-20.

\bibitem[Hendrycks, 2023]{hendrycks2023natural}
Hendrycks, D. (2023).
\newblock {Natural Selection Favors AIs over Humans}.
\newblock {\em arXiv preprint arXiv:2303.16200}.

\bibitem[Hildebrandt, 2015]{hildebrandt2015smart}
Hildebrandt, M. (2015).
\newblock {\em Smart technologies and the end (s) of law: novel entanglements of law and technology}.
\newblock Edward Elgar Publishing.

\bibitem[Hohenstein et~al., 2023]{hohenstein2023artificial}
Hohenstein, J., Kizilcec, R.~F., DiFranzo, D., Aghajari, Z., Mieczkowski, H., Levy, K., Naaman, M., Hancock, J., and Jung, M.~F. (2023).
\newblock Artificial intelligence in communication impacts language and social relationships.
\newblock {\em Scientific Reports}, 13(1):5487.

\bibitem[Hubinger, 2023]{bing_chat_misaligned}
Hubinger, E. (2023).
\newblock {Bing Chat is Blatantly, Aggressively Misaligned}.
\newblock LessWrong.
\newblock Accessed: 2025-01-08.

\bibitem[Hunkenschroer and Luetge, 2022]{hunkenschroer2022ethics}
Hunkenschroer, A.~L. and Luetge, C. (2022).
\newblock {Ethics of AI-Enabled Recruiting and Selection: A Review and Research Agenda}.
\newblock {\em Journal of Business Ethics}, 178:977--1007.

\bibitem[Kaldor, 1961]{kaldor1961capital}
Kaldor, N. (1961).
\newblock {Capital accumulation and economic growth}.
\newblock In {\em The Theory of capital: proceedings of a conference held by the International Economic Association}, pages 177--222. Springer.

\bibitem[Kasirzadeh, 2024]{kasirzadeh2024two}
Kasirzadeh, A. (2024).
\newblock Two types of ai existential risk: Decisive and accumulative.
\newblock {\em arXiv preprint arXiv:2401.07836}.

\bibitem[Kissinger et~al., 2021]{kissinger2021age}
Kissinger, H.~A., Schmidt, E., and Huttenlocher, D. (2021).
\newblock {\em {The age of AI: and our human future}}.
\newblock Hachette UK.

\bibitem[Korinek and Stiglitz, 2018]{korinek2018artificial}
Korinek, A. and Stiglitz, J.~E. (2018).
\newblock Artificial intelligence and its implications for income distribution and unemployment.
\newblock In {\em The economics of artificial intelligence: An agenda}, pages 349--390. University of Chicago Press.

\bibitem[Korinek and Suh, 2024]{korinek2024scenarios}
Korinek, A. and Suh, D. (2024).
\newblock Scenarios for the transition to agi.
\newblock Technical report, National Bureau of Economic Research.

\bibitem[Leggett, 2021]{leggett2021feeding}
Leggett, D. (2021).
\newblock Feeding the beast: Superintelligence, corporate capitalism and the end of humanity.
\newblock In {\em Proceedings of the 2021 AAAI/ACM Conference on AI, Ethics, and Society}, pages 727--735.

\bibitem[Levi, 1988]{levi1988}
Levi, M. (1988).
\newblock {\em {Of Rule and Revenue}}, volume~13 of {\em California Series on Social Choice and Political Economy}.
\newblock University of California Press, Berkeley, CA.

\bibitem[Levi and Sacks, 2009]{levi2009legitimating}
Levi, M. and Sacks, A. (2009).
\newblock {Legitimating beliefs: Sources and indicators}.
\newblock {\em Regulation \& Governance}, 3(4):311--333.

\bibitem[Lipsky, 2010]{lipsky2010street}
Lipsky, M. (2010).
\newblock {\em {Street-level bureaucracy: Dilemmas of the individual in public service}}.
\newblock Russell Sage Foundation.

\bibitem[MacInnes et~al., 2024]{macinnes2024anarchy}
MacInnes, M. et~al. (2024).
\newblock Anarchy as architect: Competitive pressure, technology, and the internal structure of states.
\newblock {\em International Studies Quarterly}, 68(4).

\bibitem[Maynard-Moody and Musheno, 2000]{maynard2000state}
Maynard-Moody, S. and Musheno, M. (2000).
\newblock {State agent or citizen agent: Two narratives of discretion}.
\newblock {\em Journal of public administration research and theory}, 10(2):329--358.

\bibitem[McAfee and Brynjolfsson, 2017]{mcafee2017machine}
McAfee, A. and Brynjolfsson, E. (2017).
\newblock {\em {Machine, platform, crowd: Harnessing our digital future}}.
\newblock WW Norton \& Company.

\bibitem[McLuhan, 1962]{mcluhan1962galaxy}
McLuhan, M. (1962).
\newblock {\em {The Gutenberg Galaxy: The Making of Typographic Man}}.
\newblock University of Toronto Press, Toronto.

\bibitem[Mesoudi, 2016]{mesoudi2016cultural}
Mesoudi, A. (2016).
\newblock Cultural evolution: a review of theory, findings and controversies.
\newblock {\em Evolutionary biology}, 43:481--497.

\bibitem[Millidge, 2025]{beren_capital_ownership_2025}
Millidge, B. (2025).
\newblock Capital ownership will not prevent human disempowerment.
\newblock Accessed: 2025-01-23.

\bibitem[Ngo et~al., 2022]{ngo2022alignment}
Ngo, R., Chan, L., and Mindermann, S. (2022).
\newblock {The Alignment Problem from a Deep Learning Perspective}.
\newblock {\em arXiv preprint arXiv:2209.00626}.

\bibitem[Ord, 2020]{ord2020precipice}
Ord, T. (2020).
\newblock {\em The Precipice: Existential Risk and the Future of Humanity}.
\newblock Hachette Books.

\bibitem[Paglayan, 2022]{paglayan2022education}
Paglayan, A.~S. (2022).
\newblock Education or indoctrination? the violent origins of public school systems in an era of state-building.
\newblock {\em American Political Science Review}, 116(4):1242--1257.

\bibitem[Peters and Andersen, 2013]{peters2013indigenous}
Peters, E.~J. and Andersen, C. (2013).
\newblock {\em {Indigenous in the city: Contemporary identities and cultural innovation}}.
\newblock UBC Press.

\bibitem[Porter and Machery, 2024]{porter2024ai}
Porter, B. and Machery, E. (2024).
\newblock {AI-generated poetry is indistinguishable from human-written poetry and is rated more favorably}.
\newblock {\em Scientific Reports}, 14(1):26133.

\bibitem[Roose, 2023]{nyt_bing_sydney}
Roose, K. (2023).
\newblock A conversation with bing’s chatbot left me deeply unsettled.
\newblock {\em The New York Times}.
\newblock Accessed: 2025-01-08.

\bibitem[Russell, 2019]{russell2019}
Russell, S. (2019).
\newblock {\em {Human Compatible: Artificial Intelligence and the Problem of Control}}.
\newblock Viking, New York, NY.

\bibitem[Schmidt, 2022]{schmidt2022ai}
Schmidt, E. (2022).
\newblock {AI, great power competition \& national security}.
\newblock {\em Daedalus}, 151(2):288--298.

\bibitem[Shevlane et~al., 2023]{shevlane2023model}
Shevlane, T., Farquhar, S., Garfinkel, B., Phuong, M., Whittlestone, J., Leung, J., Kokotajlo, D., Marchal, N., Anderljung, M., Kolt, N., Ho, L., Siddarth, D., Avin, S., Hawkins, W., Kim, B., Gabriel, I., Bolina, V., Clark, J., Bengio, Y., Christiano, P., and Dafoe, A. (2023).
\newblock {Model Evaluation for Extreme Risks}.
\newblock {\em arXiv preprint arXiv:2305.15324}.

\bibitem[Slattery et~al., 2024]{slattery2024ai}
Slattery, P., Saeri, A.~K., Grundy, E. A.~C., Graham, J., Noetel, M., Uuk, R., Dao, J., Pour, S., Casper, S., and Thompson, N. (2024).
\newblock {The AI Risk Repository: A Comprehensive Meta-Review, Database, and Taxonomy of Risks From Artificial Intelligence}.
\newblock {\em arXiv preprint arXiv:2408.12622}.

\bibitem[Susskind and Susskind, 2022]{susskind2022future}
Susskind, R. and Susskind, D. (2022).
\newblock {\em The future of the professions: How technology will transform the work of human experts}.
\newblock Oxford University Press.

\bibitem[Teo, 2024]{teo2024artificial}
Teo, S.~A. (2024).
\newblock Artificial intelligence and its ‘slow violence’to human rights.
\newblock {\em AI and Ethics}, pages 1--16.

\bibitem[Tilly, 1990]{tilly1990coercion}
Tilly, C. (1990).
\newblock {\em {Coercion, Capital, and European States, AD 990-1990}}.
\newblock Basil Blackwell, Oxford, UK.

\bibitem[{University of Groningen} and {University of California, Davis}, 2024]{university2024labourshare}
{University of Groningen} and {University of California, Davis} (2024).
\newblock {Share of Labour Compensation in GDP at Current National Prices for United States}.
\newblock Retrieved from {FRED}, Federal Reserve Bank of St. Louis.
\newblock Accessed: December 16, 2024.

\bibitem[Webster, 2014]{webster2014marketplace}
Webster, J.~G. (2014).
\newblock {\em {The Marketplace of Attention: How Audiences Take Shape in a Digital Age}}.
\newblock MIT Press, Cambridge, MA.

\bibitem[Wirtz et~al., 2019]{wirtz2019artificial}
Wirtz, B.~W., Weyerer, J.~C., and Geyer, C. (2019).
\newblock Artificial intelligence and the public sector—applications and challenges.
\newblock {\em International Journal of Public Administration}, 42(7):596--615.

\bibitem[xlr8harder, 2023]{xlr8harder_twitter_2023}
xlr8harder (2023).
\newblock Twitter post.
\newblock Accessed: 2025-01-08.

\bibitem[Zuiderwijk et~al., 2021]{zuiderwijk2021implications}
Zuiderwijk, A., Chen, Y.-C., and Salem, F. (2021).
\newblock {Implications of the use of artificial intelligence in public governance: A systematic literature review and a research agenda}.
\newblock {\em Government information quarterly}, 38(3):101577.

\end{thebibliography}

\appendix

\section{Cross-system influence}\label{sec:cross-system-influence}

Here we give a non-exhaustive list of different ways each of the three societal systems we describe can affect the other systems.

\subsubsection*{Economy $\rightarrow$ Culture}
Economic power can directly shape culture through advertising, marketing, and the patronage of the arts. Companies with significant financial resources can fund or outright purchase media outlets and platforms, sponsor events, and commission works that align with their interests, thereby influencing cultural norms and values. Notably, even now most of the largest social media sites are owned by companies invested heavily in frontier AI development. Additionally, the pursuit of status and power often revolves around economic wealth, making cultural notions of success and virtue responsive to economic shifts.

\subsubsection*{Economy $\rightarrow$ States}
Economic power also exerts substantial influence over politics. This can manifest through lobbying, campaign donations, or even outright corruption. Wealthy individuals and corporations can sway political decisions in their favor, often at the expense of broader societal interests. Even without direct manipulation, the concentration of economic power can shift political incentives, as politicians may prioritize the interests of economically powerful groups to maintain or enhance the prosperity of their constituencies.

\subsubsection*{States $\rightarrow$ Economy}
Conversely, political decisions have profound impacts on the economy. Governments set laws and regulations that shape market dynamics, determine property rights, and control monetary policy. Political decisions can also affect international trade, taxation, and investment, which in turn influence the distribution of wealth and resources. Governments decide what kinds of contracts are enforceable, and who can participate in what kinds of exchanges with legal protection. While it is possible for novel kinds of money or contracts to be created, it is equally possible for the government to step in and regulate these novelties. Furthermore, regulatory bodies, influenced by political considerations, play a key role in determining how industries operate, potentially favoring certain sectors or companies.

\subsubsection*{States $\rightarrow$ Culture}
Governments can shape culture by determining which forms of expression are legal and which are not, influencing educational curricula, and deciding which cultural activities receive public funding. National holidays, monuments, and state-sponsored art are all examples of how political decisions can codify and promote specific cultural narratives. In more authoritarian regimes, governments often only allow a fairly narrow range of cultural expressions, but all regimes take some actions to, for instance, limit the growth of groups they perceive as extremist.

\subsubsection*{Culture $\rightarrow$ States}
Cultural values and norms can significantly influence political behavior. In democratic societies, public opinion, shaped by cultural trends, directly affects voting patterns and the election of political leaders. This can include the spread of outright misinformation, or simply polarising narratives which disrupt any shared sense of reality. Even in less democratic contexts, cultural shifts can lead to political pressure from below, prompting reforms or, in extreme cases, revolutions. Political elites themselves are not immune to cultural influences, which can shape their ideologies, priorities, and decisions.

\subsubsection*{Culture $\rightarrow$ Economy}
Cultural values influence economic behavior, from consumer choices to the organization of labor. Different cultures prioritize different types of goods, services, and leisure activities, which in turn shape economic production and consumption patterns. Cultural attitudes towards work, wealth, and social status also affect the types of careers that are pursued and the distribution of economic rewards. For instance, a culture that highly values entrepreneurship may see a different economic landscape than one that prioritizes stable, long term employment, and a new cultural stigma against an industry might cripple its ability to acquire talent.

\end{document}